\documentclass[journal]{IEEEtran}
\usepackage[dvips]{graphics}
\usepackage{graphicx}
\usepackage{times}
\usepackage{amsmath}
\usepackage{amssymb}
\usepackage{fancybox}
\usepackage{color}
\newtheorem{lemma}{\bf Lemma}
\newtheorem{theorem}{\bf Theorem}

\newcommand{\bee}{\begin{eqnarray}}
\newcommand{\eee}{\end{eqnarray}}
\newcommand{\be}{\begin{equation}}
\newcommand{\ee}{\end{equation}}

\newcommand{\al}[1]{\begin{align} #1 \end{align}}

\newcommand{\equ}[1]{\begin{equation} #1 \end{equation}}

\newcommand{\mb}{\mathbf}

\newcommand{\nnb}{\nonumber}

\newcommand{\qc}{{\bf c}}
\newcommand{\qe}{{\bf e}}

\newcommand{\qn}{{\bf n}}

\newcommand{\qu}{{\bf u}}

\newcommand{\qx}{{\bf x}}
\newcommand{\qy}{{\bf y}}

\newcommand{\qA}{{\bf A}}
\newcommand{\qB}{{\bf B}}

\newcommand{\qG}{{\bf G}}
\newcommand{\qH}{{\bf H}}
\newcommand{\qI}{{\bf I}}

\newcommand{\qQ}{{\bf Q}}

\newcommand{\qS}{{\bf S}}

\begin{document}
%
\title{Closed Form Secrecy Capacity of MIMO Wiretap Channels with Two Transmit Antennas}

\author{\IEEEauthorblockN{Jiangyuan Li and Athina Petropulu}\\
\IEEEauthorblockA{Department of Electrical and Computer Engineering\\
Rutgers-The State University of New Jersey, New Brunswick, NJ 08854}
}

\maketitle

\begin{abstract}
\footnote{Work supported by the Office of Naval Research under grant ONR-N-00010710500 and the
National Science Foundation under grant CNS-0905425.}
A Gaussian multiple-input multiple-output (MIMO) wiretap channel model is considered. The input is a two-antenna transmitter, while the outputs are
the legitimate receiver and an eavesdropper, both equipped with multiple antennas. All channels are assumed to be known.
The problem of obtaining the optimal input covariance matrix that achieves secrecy capacity subject to a power constraint is addressed, and a
closed-form expression for the secrecy capacity is obtained.

\end{abstract}

\begin{IEEEkeywords}
Secrecy capacity, MIMO wiretap channel, physical layer security
\end{IEEEkeywords}


\section{Introduction}

Wireless physical layer based security
approaches present an alternative to cryptographic approaches.
They enhance the security of a communication system by exploiting the physical characteristics of the
wireless channel.
Wyner identified the secrecy capacity of the single user memoryless wiretap channel \cite{Wyner}.
Later, the Gaussian scalar wiretap channel was studied in \cite{Hellman}.
More recently, the secrecy capacity of the MIMO wiretap channel under power constraints was analyzed in
\cite{Khisti2} and \cite{Khisti3}, while the same problem  under power-covariance constraints was studied in
\cite{Shamai} and \cite{Poor}.
In \cite{Shafiee}, the case $(n_T, n_R, n_E)=(2,2,1)$ (two transmit antennas, two receive antennas and one eavesdropper antenna) was analyzed, and it was shown that under certain assumptions on the channels, beamforming is optimal.
In \cite{Hassibi-Allerton}, the  $(2,2,2)$ case was studied under equality power constraint and the positive definiteness assumption
$\qH_E^\dagger\qH_E\succ 0$.
However, the solution was not given, and moreover, it does not show that equality power constraint
is equivalent to the more common inequality power constraint.

In this letter, we study the secrecy capacity for a MIMO wiretap channel with two transmit antennas and an arbitrary number of receive and eavesdropper antennas.
The main result is a closed form expression for the capacity, that is obtained based on the roots of a quadratic and a quartic equation.

\smallskip

{\em Notation -}
Upper case and lower case bold symbols denote matrices and vectors, respectively.
Superscripts $\ast$, $T$ and $\dagger$
denote respectively conjugate, transposition and conjugate transposition. $\mathrm{det}(\qA)$ and
$\mathrm{Tr}({\mb A})$ denote the determinant and trace of the matrix $\mb A$, respectively.
$\lambda_{\max}(\qA)$ denotes the largest eigenvalue of the matrix $\qA$.
${\mb A}\succeq 0$ denotes that the matrix $\qA$ is Hermitian positive semi-definite, and $\qA\succ 0$ denotes that the matrix $\qA$
is Hermitian positive definite.
$|a|$ denotes the absolute value of the complex number $a$.
$\qI_n$ denotes the identity matrix of order $n$ (the subscript is dropped when the dimension is obvious).

\section{System Model and Problem Statement}\label{sec:2}

Consider a MIMO wiretap channel, where the transmitter is equipped with $n_T=2$ antennas, while
the legitimate receiver and an eavesdropper have $n_R$ and $n_E$ antennas, respectively.
The received signals at the legitimate receiver and the eavesdropper are respectively given by
\equ{
\qy_R = \qH_R\qx+\qn_R, \ \mathrm{and}\ \qy_E = \qH_E\qx+\qn_E
}
where $\qH_R$ ($n_R\times 2$), $\qH_E$ ($n_E\times 2$) are respectively matrices representing the channel between transmitter and legitimate receiver,
and  transmitter and eavesdropper;
$\qx$ is the $2 \times 1$ transmitted signal vector with zero mean and $2 \times 2$ covariance matrix $P\qQ$,
where $P$ is the power constraint, $\qQ\succeq 0$ and $\mathrm{Tr}(\qQ)\le 1$;
$\qn_R$ and $\qn_E$ are Gaussian noise vectors with zero mean and covariance matrices
$\sigma^2\qI_{n_R}$ and $\sigma^2\qI_{n_E}$, respectively.
We consider the scenario in which the transmitter has perfect channel state information (CSI) on $\qH_R$ and $\qH_E$.

The secrecy capacity for this scenario is  \cite{Khisti3}
\equ{
C_s \triangleq \max_{\qQ\succeq 0, \ \mathrm{Tr}(\qQ)\le 1}\ C_s(\qQ)\label{SecRateMax1}
}
where $C_s(\qQ) = \log\mathrm{det}(\qI+\qQ\qS_R) -\log\mathrm{det}(\qI+\qQ\qS_E)$
and $\qS_R=\rho\qH_R^\dagger\qH_R$, $\qS_E=\rho\qH_E^\dagger\qH_E$, $\rho = P/\sigma^2$.

\smallskip

\begin{lemma}\label{lem:1}
{\em The sufficient and necessary condition for $C_s>0$ is that $\qH_R^\dagger\qH_R-\qH_E^\dagger\qH_E$
has at least a positive eigenvalue.}
\end{lemma}
The proof is simple. Please refer to \cite[Lemma 1]{Li} for details.

\smallskip

In this paper, we assume that the condition in Lemma \ref{lem:1} holds which ensures $C_s>0$ and hence $\qQ^\star \ne 0$.

\section{Closed Form Secrecy Capacity}

The main result of this paper is contained in Theorem \ref{Theo:Cap}.

\medskip

\begin{theorem}\label{Theo:Cap}
{\em Let $\tau_1$ be the largest real root (if any) of the quadratic equation
\equ{
-\tau^2 q_3 + \tau p_3 - q_6 = 0. \label{Quadratic}
}
Let $\tau_2$ be the largest real root (if any) of the quartic equation
\al{
&(-\tau^2 q_2 + \tau p_2 - q_5)^2 \nnb\\
&\quad\quad - 4(-\tau^2 q_1 + \tau p_1 - q_4)(-\tau^2 q_3 + \tau p_3 - q_6) = 0 \label{Quartic}
}
such that
\equ{
0< -\frac{-\tau_2^2 q_2 + \tau_2 p_2 - q_5}{2(-\tau_2^2 q_1 + \tau_2 p_1 - q_4)} < 1.
}
Then the secrecy capacity equals
\equ{
C_s = \log \big(\max\{\tau_1, \tau_2\}\big). \label{Cstau}
}
}
\end{theorem}
The coefficients $\{p_i\}$, $\{q_j\}$ are given by
\al{
p_1 &= -b_1^\ast b_2 - b_1 b_2^\ast  - (1+a_1)(a_2c_2-|b_2|^2) \nnb\\
&\qquad\quad -(1+a_2)(a_1c_1-|b_1|^2), \nnb\\
p_2 &= 2b_1^\ast b_2 + 2b_1b_2^\ast+(1+a_1)(a_2-c_2 + a_2c_2-|b_2|^2)\nnb\\
&\quad\quad\quad +(1+a_2)(a_1-c_1 + a_1c_1-|b_1|^2), \nnb\\
p_3 &= (1+a_1)(1+c_2)+(1+a_2)(1+c_1)-b_1^\ast b_2-b_1b_2^\ast, \nnb\\
q_1 &= -a_2(c_2 + a_2c_2-|b_2|^2), \nnb\\
q_2 &= a_2-c_2+a_2^2+|b_2|^2+a_2(a_2c_2-|b_2|^2), \nnb\\
q_3 &= 1+a_2+c_2 + a_2c_2-|b_2|^2, \nnb\\
q_4 &= -a_1(c_1 + a_1c_1-|b_1|^2), \nnb\\
q_5 &= a_1-c_1+a_1^2+|b_1|^2+a_1(a_1c_1-|b_1|^2), \nnb\\
q_6 &= 1+a_1+c_1 + a_1c_1-|b_1|^2      \label{piqi}
}
where $a_1, b_1, c_1, a_2, b_2, c_2$ are entries of $\qS_R$ and $\qS_E$, i.e.,
\equ{
\qS_R = \left(
          \begin{array}{cc}
            a_1 & b_1 \\
            b_1^\ast & c_1 \\
          \end{array}
        \right), \ \mathrm{and} \ \qS_E = \left(
                                            \begin{array}{cc}
                                              a_2 & b_2 \\
                                              b_2^\ast & c_2 \\
                                            \end{array}
                                          \right).
}
{\em Remarks}: The quartic equation (\ref{Quartic}) can be solved by radicals (closed form) \cite[p. 87]{King}.

\medskip

{\bf Proof of Theorem \ref{Theo:Cap}}:

\medskip

For the proof we need the following two lemmas.

\smallskip

\begin{lemma}\label{Lem:2-by-2Mat}
{\em Let $\qe_1=[1\ 0]^T$. Any $2\times 2$ matrix $\qQ \ne 0$ with $\qQ\succeq 0$ and $\mathrm{Tr}(\qQ)\le 1$ can be expressed as
\equ{
\qQ = x\qe_1\qe_1^\dagger + (1-x)\qu\qu^\dagger
}
where $x$ is a real number with $0\le x < 1$ and $\qu\ne 0$ is a $2\times 1$ vector with $\qu^\dagger\qu \le 1$.
}
\end{lemma}
{\em Proof}: One can write $\qQ=x\qe_1\qe_1^\dagger + (\qQ - x\qe_1\qe_1^\dagger)$ where $x$ is chosen to satisfy $\det(\qQ - x\qe_1\qe_1^\dagger)=0$.

\medskip

\begin{lemma}\label{Lem:det}
{\em For vectors $\qc_1$, $\qc_2$, $\qc_3$ and $\qc_4$, it holds that
\equ{
\det(\qI + \qc_1\qc_2^\dagger + \qc_3\qc_4^\dagger) = (1+\qc_2^\dagger\qc_1)(1+\qc_4^\dagger\qc_3)-\qc_4^\dagger\qc_1\qc_2^\dagger\qc_3.
}
}
\end{lemma}
{\em Proof}:
Using $\det(\qI+\qA\qB)=\det(\qI+\qB\qA)$ \cite[p. 420]{Harville}, it holds that
$\det(\qI + \qc_1\qc_2^\dagger + \qc_3\qc_4^\dagger)=\det(\qI+[\qc_1 \ \qc_3][\qc_2 \ \qc_4]^\dagger)
=\det(\qI_2+[\qc_2 \ \qc_4]^\dagger[\qc_1 \ \qc_3])$ which leads to the desired result.

\medskip

Since $C_s>0$, it holds that $\qQ^\star \ne 0$. From Lemma \ref{Lem:2-by-2Mat},
we let $\qQ=x\qe_1\qe_1^\dagger + (1-x)\qu\qu^\dagger$ with $0\le x < 1$, $\qu^\dagger\qu\le 1$ and use Lemma \ref{Lem:det} to rewrite
$C_s(\qQ)$ as
\al{
&C_s(x, \qu) = \nnb\\
&\log\!\frac
{(1 \!\! + \! x\qe_1^\dagger\qS_R\qe_1)(1 \!\! + \! (1 \! - \! x)\qu^\dagger\qS_R\qu) \! - \! (x \! - \! x^2)|\qu^\dagger\qS_R\qe_1|^2}
{(1 \!\! + \! x\qe_1^\dagger\qS_E\qe_1)(1 \!\! + \! (1 \! - \! x)\qu^\dagger\qS_E\qu) \! - \! (x \! - \! x^2)|\qu^\dagger\qS_E\qe_1|^2}
\nnb\\
&= \log \frac{1 \!\! + \! x\qe_1^\dagger\qS_R\qe_1 \!\! + \! (1 \! - \! x)\qu^\dagger\big(\qS_R \!\! + \! x\det(\qS_R)\qe_2\qe_2^\dagger\big)\qu}
{1 \!\! + \! x\qe_1^\dagger\qS_E\qe_1 \!\! + \! (1 \! - \! x)\qu^\dagger\big(\qS_E \!\! + \! x\det(\qS_E)\qe_2\qe_2^\dagger\big)\qu}
\label{Cs-x-u}
}
where $\qe_2=[0\ 1]^T$. Here, to obtain (\ref{Cs-x-u}), we have used the identity $(\qe_1^\dagger\qS_R\qe_1)\qS_R - \qS_R\qe_1\qe_1^\dagger\qS_R = \det(\qS_R)\qe_2\qe_2^\dagger$.
Since $C_s>0$, there exists a solution $\qu^\star$ with ${\qu^\star}^\dagger{\qu^\star}=1$. To see why, assume any solution $\qu^\star$ satisfies ${\qu^\star}^\dagger{\qu^\star}<1$.
It is easy to verify that $C_s(x^\star, \sqrt{t}\ \qu^\star)$ is a monotonic function of $t \in [0, 1/({\qu^\star}^\dagger{\qu^\star})]$.
Thus, either $t=0$ or $t=1/({\qu^\star}^\dagger{\qu^\star})$ achieves a larger objective value than $t=1$.
But this contradicts the optimality of $\qu^\star$.
From this result, we write
\equ{
C_s(x, \qu) =
\log\frac{\qu^\dagger\qG_1(x)\qu}
{\qu^\dagger\qG_2(x)\qu} \label{Cs_x_u}
}
where $\qG_1(x) = (1+x\qe_1^\dagger\qS_R\qe_1)\qI+(1-x)(\qS_R + x\det(\qS_R)\qe_2\qe_2^\dagger)$ and
$\qG_2(x) = (1+x\qe_1^\dagger\qS_E\qe_1)\qI+(1-x)(\qS_E + x\det(\qS_E)\qe_2\qe_2^\dagger)$.
For fixed $x$, the optimal $\qu$ is the unit-norm eigenvector associated with the largest eigenvalue
of the matrix $\qG_2(x)^{-1}\qG_1(x)$.
The problem becomes
\al{
&\max_{x}\ \big[C_s(x)=\log\lambda_{\max}\big(\qG_2(x)^{-1}\qG_1(x)\big)\big]\label{2by2Capa-1}\\
&\mathrm{s.t.}\quad 0\le x < 1. \nnb
}
By using the fact that for a $2\times 2$ matrix $\qB$,
\equ{
\lambda_{\max}(\qB)=\frac{\mathrm{Tr}(\qB)+\sqrt{(\mathrm{Tr}(\qB))^2-4\det(\qB)}}{2}, \label{max-eig}
}
we get (simply finding $\mathrm{Tr}(\qG_2(x)^{-1}\qG_1(x))=f_1(x)/f_2(x)$ and $\det(\qG_2(x)^{-1}\qG_1(x))=f_3(x)/f_2(x)$)
\equ{
\lambda_{\max}(\qG_2(x)^{\! - \! 1}\qG_1(x)) = \frac{f_1(x) + \! \sqrt{(f_1(x))^2 \! - \! 4f_2(x)f_3(x)}}{2f_2(x)}
}
where $f_1(x) = p_1x^2+p_2x+p_3$, $f_2(x) = q_1x^2+q_2x+q_3$, $f_3(x) = q_4x^2+q_5x+q_6$,
and the coefficients $p_i$'s and $q_i$'s are given in (\ref{piqi}).
Since $\qG_1(x)\succ 0$ and $\qG_2(x)\succ 0$, it holds that that $f_i(x)>0$ for $0\le x < 1$, $i=1,2,3$.

From the above result, the problem of (\ref{2by2Capa-1}) becomes
\al{
&\max_{x, \ \tau}\quad \log\tau \label{2by2Capa-2}\\
&\mathrm{s.t.}\ \ 0\le x < 1, \ \ \frac{f_1(x) + \! \sqrt{(f_1(x))^2 \! - \! 4f_2(x)f_3(x)}}{2f_2(x)} \ge \tau \nnb
}
which is equivalent to
\al{
&\max_{x, \ \tau}\quad \log\tau \label{2by2Capa-3}\\
&\mathrm{s.t.}\quad 0\le x < 1, \ \mathrm{and}\ -\tau^2 f_2(x)+\tau f_1(x)-f_3(x) \ge 0. \nnb
}
The equivalence of (\ref{2by2Capa-2}) and (\ref{2by2Capa-3}) can be verified as follows.
Firstly, the second constraint in (\ref{2by2Capa-3}) can be rewritten as
\equ{
(f_1(x))^2-4f_2(x)f_3(x) \ge (2f_2(x)\tau - f_1(x))^2. \label{2ndConstraint}
}
For the optimal $\tau$ and $x$, it holds that $2f_2(x)\tau - f_1(x)\ge 0$. Otherwise,
one can choose $\tau'=f_1(x)/f_2(x) - \tau > \tau$ such that $(\tau', x)$ satisfies the constraint (\ref{2ndConstraint}). But this contradicts the optimality of $\tau$.
With this fact, (\ref{2ndConstraint}) leads to the second constraint in (\ref{2by2Capa-2}).
Secondly, for the optimal $\tau$ and $x$, the second constraint in (\ref{2by2Capa-2}) holds with equality which leads to $-\tau^2 f_2(x)+\tau f_1(x)-f_3(x) = 0$.
The desired result follows.

\smallskip

Next we solve the problem of (\ref{2by2Capa-3}).
Denote $F(x)=-\tau^2 f_2(x)+\tau f_1(x)-f_3(x)$ which can be rewritten as
\equ{
F(x) = A_1x^2 + B_1x + C_1,
}
where $A_1 = -\tau^2 q_1 + \tau p_1 - q_4$, $B_1 = -\tau^2 q_2 + \tau p_2 - q_5$,
and $C_1 = -\tau^2 q_3 + \tau p_3 - q_6$.
One want to find the maximal $\tau$ (denoted as $\tau^\star$) such that
there exists at least a $x\in [0, 1)$ satisfying $F(x)\ge 0$.
In other words, for $\tau > \tau^\star$, there exists no $x\in [0, 1)$ such that
$F(x) \ge 0$.
Since $F(x)$ is a quadratic function, this fact leads to two possible situations:
\begin{itemize}
  \item [1)] The optimal $x$ (denoted as $x^\star$)
satisfies $0 < x^\star < 1$,
then $A_1\ne 0$ and $x^\star$ is the repeated root of $F(x)=0$, i.e.,
\equ{
B_1^2-4A_1C_1=0, \ \mathrm{and}\ 0< x^\star = -\frac{B_1}{2A_1}< 1.
}
In fact, if $x^\star$ is not the repeated root of $F(x)=0$, then one can find
$\tau = \tau^\star + \epsilon$ with an enough small $\epsilon > 0$ such that there exists
a $x\in [0, 1)$ satisfying $F(x)\ge 0$.
This is because the two different real roots of a quadratic equation both depend continuously on its coefficients.
But this contradicts the optimality of $\tau^\star$.
  \item [2)] $x^\star=0$ which leads to $C_1=0$.
\end{itemize}
From the above analysis, one can find $\tau^\star$ by solving $C_1=0$ and $B_1^2-4A_1C_1=0$.
This completes the proof.

\section{Numerical Simulations}\label{sec:sim}

\begin{figure}[hbtp]
\centering
\includegraphics[width=3.2in]{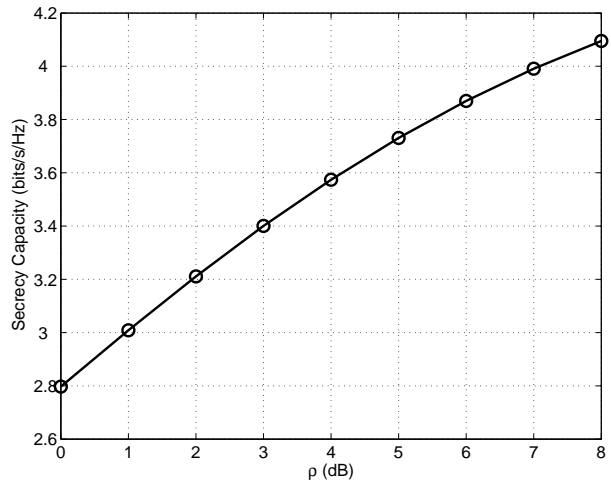}
\caption{Secrecy capacity for different $\rho$.}
\label{fig:CsSNR}
\end{figure}

First, we consider a MIMO wiretap channel in which $n_T=2$, $n_R=n_E=3$ and
\al{
\qH_R &= {\small
          \left(
          \begin{array}{rr}
            0.7442 + 1.4223\mathrm{i} & 1.1740 - 1.8109\mathrm{i} \\
            -0.5172 + 0.4116\mathrm{i} & -1.3020 + 0.2417\mathrm{i} \\
            1.9755 + 0.4169\mathrm{i} & -0.7105 + 0.7272\mathrm{i} \\
          \end{array}
        \right)
        }, \nnb\\
\qH_E &= {\small
           \left(
           \begin{array}{rr}
             -0.4503 + 0.9711\mathrm{i} & -0.7453 + 1.1555\mathrm{i} \\
             -0.7089 + 0.1272\mathrm{i} & -0.0506 + 0.5835\mathrm{i} \\
             -0.1313 - 0.3833\mathrm{i} & 0.1974 + 0.1632\mathrm{i} \\
           \end{array}
         \right)
         }. \nnb
}
We set $\rho=5\, \mathrm{dB}$.
The quadratic equation (\ref{Quadratic}) has two real roots
$(1.4247, 10.8607)$, and hence $\tau_1 = 10.8607$.
The quartic equation (\ref{Quartic}) has
four roots $19.0710, 13.2768, 3.4529 \pm 1.5230 \times 10^{-8}\mathrm{i}$,
and $\tau_2 = 13.2768$ with $x = -B_1/(2A_1) = 0.3189$.
Thus, the secrecy capacity is
\equ{
C_s = \log(\tau_2)/\log(2) = 3.7308 \, \mathrm{(bits/s/Hz)} \nnb
}
which is achieved at $x^\star = 0.3189$.
The optimal input covariance matrix is
\equ{
\qQ^\star = \left(
              \begin{array}{ll}
                0.5435 & -0.3198 + 0.0164\mathrm{i} \\
                -0.3198 - 0.0164\mathrm{i} & 0.4565 \\
              \end{array}
            \right) \nnb
}
which has rank two.
Fig. \ref{fig:CsSNR} plots the secrecy capacity for different $\rho$.

Second, we change the previous example to $n_E=1$ and $\qH_E=[ -1.2480 - 0.2893\mathrm{i}, 4.6312 + 0.2417\mathrm{i}]$.
It holds that $\qH_E(\qH_R^\dagger \qH_R)^{-1}\qH_E^\dagger > 1$ (this is exactly the condition in \cite[Lemma 1]{Shafiee}).
The quadratic equation (\ref{Quadratic}) has two real roots $0.3453, 20.5293$ and hence $\tau_1=20.5293$.
The quartic equation (\ref{Quartic}) has four roots $6.8179 \pm 2.2258\mathrm{i}$ and
$3.7837 \pm 4.38 \times 10^{-8}\mathrm{i}$. Thus, $x^\star=0$ and hence beamforming is optimal.
This is consistent with the result in \cite[Lemma 1]{Shafiee}).

\section{Conclusion}\label{sec:conclusion}

We have studied a Gaussian MIMO wiretap channel in which there exists a transmitter with two antennas, a legitimate receiver and an eavesdropper both equipped with multiple antennas.
We derived the the secrecy capacity in closed form.

\end{document}